\documentclass[12pt,letterpaper]{article}
\usepackage[a4paper, total={7in, 10in}]{geometry}

\usepackage{helvet}
\usepackage{authblk}
\usepackage{hyperref}

\usepackage{makecell}
\usepackage{multirow}
\usepackage{xcolor}
\usepackage{graphicx} 

\usepackage[left]{lineno}
\usepackage{pdfpages}

\makeatletter
\renewcommand{\maketitle}{\bgroup\setlength{\parindent}{0pt}
\begin{flushleft}
  \textbf{\@title}
  
  \@author
\end{flushleft}\egroup}
\makeatother

\title{AI-enhanced collective intelligence} %AI-enhanced Collective Intelligence: The State of the Art and Prospects
\date{}

\author[1,2]{Hao Cui}
\author[1,2,3]{Taha Yasseri}
\affil[1]{School of Sociology, University College Dublin, Dublin, Ireland}
\affil[2]{Geary Institute for Public Policy, University College Dublin, Dublin, Ireland}
\affil[3]{School of Social Sciences and Philosophy, Trinity College Dublin, Dublin, Ireland}
\affil[*]{taha.yasseri@tcd.ie}

\usepackage[super,comma,sort&compress]{natbib}

\begin{document}

\maketitle

\section*{Summary}

%The summary (abstract) should consist of a single paragraph of 150 words or fewer.

Current societal challenges exceed the capacity of humans operating either alone or collectively. As AI evolves, its role within human collectives will vary from an assistive tool to a participatory member. Humans and AI possess complementary capabilities that, together, can surpass the collective intelligence of either humans or AI in isolation. However, the interactions in human-AI systems are inherently complex, involving intricate processes and interdependencies. This review incorporates perspectives from complex network science to conceptualize a multilayer representation of human-AI collective intelligence, comprising cognition, physical, and information layers. Within this multilayer network, humans and AI agents exhibit varying characteristics; humans differ in diversity from surface-level to deep-level attributes, while AI agents range in degrees of functionality and anthropomorphism. We explore how agents' diversity and interactions influence the system's collective intelligence and analyze real-world instances of AI-enhanced collective intelligence. We conclude by considering potential challenges and future developments in this field.

%The current societal challenges exceed the capacity of human individual or collective effort alone. As AI evolves, its role within human collectives is poised to vary from an assistive tool to a participatory member. Humans and AI possess complementary capabilities that, when synergized, can achieve a level of collective intelligence that surpasses the collective capabilities of either humans or AI in isolation. However, the interactions in human-AI systems are inherently complex, involving intricate processes and interdependencies. This narrative review incorporates perspectives from complex network science to conceptualize a multilayer representation of human-AI collective intelligence, comprising cognition, physical, and information layers. Within this multilayer network, humans and AI agents exhibit varying characteristics; humans differ in diversity from surface-level to deep-level attributes, while AI agents range in degrees of functionality and anthropomorphism. The interplay among these agents shapes the overall structure and dynamics of the system. We explore how agents' diversity and interactions influence the system's collective intelligence. Furthermore, we present an analysis of real-world instances of AI-enhanced collective intelligence. We conclude by addressing the potential challenges in AI-enhanced collective intelligence and offer perspectives on future developments in this field.

\section*{Keywords}

AI, Collective Intelligence, Hybrid Intelligence, Multi-agent systems, Human-machine networks, Human-machine Intelligence

%Include up to 10 keywords, separated by commas. (Keywords are not carried over from Editorial Manager. Only keywords included in the main text will be used in the final article metadata.)


%\maketitle

\section*{Introduction}
Our societies form, sustain, and function because of our intelligence. In the animal kingdom, the more intelligent the animals are, the more social they are. Dunbar’s ``social brain theory"~\cite{dunbar1998social} even suggests that significantly more superior human intelligence compared with other primates is the result of our need to be able to manage and maintain our social lives; it is humans’ outstanding sociality that forced them to become more intelligent. The bottom line is that in human societies, collectives are intelligent, and intelligence is collective. As a concept, Collective Intelligence (CI) refers to the emergent outcome of the collective efforts of many individuals. The superiority of collective intelligence to the intelligence of any of the individuals who contributed to it has been demonstrated scientifically and practically in numerous instances~\cite{bonabeau2009decisions}.

In every new chapter in the history of Information and Communication Technologies (ICT), collective intelligence has been elevated to a higher level through more readily and cheaply available platforms for collaboration and exchanging ideas. The Internet, especially the World Wide Web, was designed to foster large-scale collaboration. Tim Berners-Lee's initial intention in developing HTTP, which later led to the development of the WWW, was to facilitate collaboration between CERN researchers. Some thirty years later, the Web facilitates the largest collaborative projects in human history, such as Wikipedia \cite{yasseri2013value}, citizen science projects \cite{ibrahim2021gender}, collaborative software development, and collaborative tagging projects \cite{Yasseri_and_Menczer_2023}, all exemplifying technology-enhanced collective intelligence %***** 

Even though large-scale knowledge-generating collaborations, crowdsourcing, and, in a more general form, collective intelligence are not Internet phenomena per se, we can see this new technology has undoubtedly boosted the state-of-the-art in collaborative knowledge creation, task execution, and collective decision-making. Artificial Intelligence (AI), similar to the Web and Internet-based technologies, is reshaping different aspects of our lives at a tremendous pace. In light of this rapidly evolving situation, it is essential to consider how, similar to previous disruptions by new ICTs, AI can magnify and enhance our collective intelligence.

\subsection*{Collective intelligence} 

Intelligent agents (humans, animals, and intelligent artifacts) often interact with others to tackle complex problems. They combine their knowledge and personal information with information acquired through interaction and social information, resulting in superior collective intelligence. The concept of collective intelligence has been redefined over the years, primarily due to different research streams that have employed the term in qualitatively different ways~\cite{woolley2024understanding}. 

\begin{table}[h!]
    \centering
    \begin{tabular}{ m{3cm}  m{4.5cm}  m{6.5cm} }
        \hline
        \textbf{Aspect} & \textbf{Wisdom of Crowds} & \textbf{Collective Intelligence} \\ 
        \hline
        \textbf{Interaction} & Minimal & High, involving significant collaboration and coordination. \\
        \hline
       % \textbf{Integration} & Aggregation of independent inputs & Emergent and adaptable integration from coordinated efforts. \\
       % \hline
        \textbf{Dependence} & Independent & Interdependent and adaptable processes to changing conditions. \\
        \hline
        \textbf{Mechanism} & Aggregation & Emergent and adaptable integration from coordinated efforts.\\ %Dynamic processes for continuous emergence and adaptation. \\
        \hline
%        \textbf{Examples} & Market predictions, crowdsourcing & Research teams, adaptive AI systems, resilient organizations. \\
%        \hline
%        \textbf{Key Characteristics} & Diversity, independence, aggregation & Collaboration, adaptability, ability to solve complex issues. \\
%        \hline
%        \textbf{Focus} & Immediate collective judgment & Long-term problem-solving, evolving to meet complex needs. \\
        \textbf{Nature} & Static, immediate collective judgment %(one-time aggregation) 
        & Dynamic and adaptive, evolving to meet complex needs. \\ %, designed to evolve over time. \\
        \hline
    \end{tabular}
    \caption{Comparison between Wisdom of Crowds and Collective Intelligence}
    \label{tab:comparison}
\end{table}

One stream of research uses collective intelligence to refer to the outcome of collaboration produced by amalgamation of the input from a large crowd, exemplified in scenarios such as online contests or crowdsourced science~\cite{aristeidou2020online, kittur2009herding, malone2018superminds}.  
Here, the concept of collective intelligence is often mixed up with the \emph{wisdom of crowds (WoC)}. 
Both rely on the idea that collective input from a diverse group typically yields better outcomes than any individual member's.
The wisdom of crowds involves aggregating independent judgments from a large, diverse group to arrive at collective decisions or predictions, often through statistical averaging~\cite{surowiecki2004wisdom}. 
In contrast, collective intelligence pertains to the achievements of collaborating groups, whose sizes can vary. It emerges from the synergy of interactions and mutual feedback among intelligent agents~\cite{grasso2012collective}  working towards a common goal through interconnected efforts. 
However, social information and interconnection have also been studied in the context of WoC~\cite{lorenz2011social, becker2017network}. Despite the conflict in the literature on the boundaries and overlaps between the two, almost all the AI enhancements we discuss in this paper in the context of CI can also be implemented in WoC scenarios~\cite{kurvers2023automating}. Nevertheless, for the sake of simplicity, we only use the term collective intelligence for the rest of the paper. Table.~\ref{tab:comparison} compares key aspects of the two concepts.

Another stream of research defines collective intelligence as an ability that ``can be designed to solve a wide range of problems over time in the face of complex and changing environmental conditions"~\cite{woolley2024understanding}. Converging evidence suggests the presence of a general collective intelligence factor (``c-factor") that serves as a predictor for a group’s performance across a wide range of tasks ~\cite{woolley2010evidence, woolley2015collective}, similar to the individual's general intelligence (g-factor) but extended to groups. Research has found that this ``c-factor" is not strongly correlated with the average or maximum individual intelligence of the group members, suggesting that collective intelligence is more than just the sum of the individual intelligence present in the group~\cite{woolley2010evidence}. In this sense, this definition is virtually compatible with the one above. 

A group's collective intelligence is fundamentally influenced by its composition and interactions. Past work indicates a positive correlation between the ``c-factor" and factors such as the average social sensitivity ~\cite{woolley2010evidence}, the equitable distribution of conversational participation ~\cite{woolley2010evidence}, and the percentage of female members in the group ~\cite{woolley2010evidence}. Groups with diverse perspectives and skill sets are likelier to cultivate innovative solutions that might not emerge in a more homogenous group. Effective interaction processes, encompassing coordination, cooperation~\cite{malone2022handbook}, and communication patterns~\cite{engel2015collective} are crucial. Prior studies have identified three distinct socio-cognitive transactive systems ~\cite{woolley2023using} responsible for managing collective memory \cite{yasseri2022collective}, collective attention \cite{gildersleve2023between}, and collective reasoning \cite{wyss2017artificial}, all of which are essential for the emergence of collective intelligence.

\subsection*{Artificial intelligence}
The definition of artificial intelligence has undergone numerous evolutions in the past semi-century. Artificial intelligence is the simulation of human intelligence processes by machines \cite{konar2018artificial}, especially computer systems. 
Previous research considers the classification of AI in distinct ways~\cite{hassani2020artificial}. 
The first classification categorizes AI based on its human-like cognitive abilities, such as 
thinking and feeling, and contains four primary AI
types~\cite{hassani2020artificial}: reactive AI, limited memory AI, theory of mind AI, and self-aware AI. Limited-memory AI is the most common at the current time, ranging from virtual assistants to chatbots to self-driving vehicles. Such AI can learn from historical data to recognize patterns, generate new knowledge and understanding, and inform subsequent decisions. 

Yet, another classification is a technology-oriented approach that categorizes AI into Artificial Narrow Intelligence (ANI), Artificial General Intelligence (AGI), and Artificial Superintelligence (ASI). 
ANI involves creating computer systems capable of performing specific tasks like human intelligence~\cite{rawat2023artificial}, but often surpass humans in efficiency and accuracy~\cite{kuusi2022scenarios}. 
ANI machines have a narrow range of capabilities and can represent all existing AI. AGI refers to machines that exhibit the human ability to learn, perceive, and understand a wide range of intellectual tasks~\cite{goertzel2014artificial, MITTechReview2023DeepMindAGI}. Finally, ASI's primary goal is to develop a machine with cognitive abilities higher than those of humans~\cite{hassani2020artificial}. 

Existing AI can perform tasks including but not limited to complex calculations, language translation, facial recognition, and financial market prediction. Recent developments in generative AI, such as OpenAI's text-to-video model Sora  (openai.com/sora), exemplify AI's potential in creative industries. As AI technology evolves, it is poised to bring groundbreaking advancements across various fields. Beyond content creation, as AI deepens the understanding of the physical world and develops its simulation ability at the perceptual and cognitive level, it is possible to develop more super-intelligent tools in various fields.

Current AI systems can process vast amounts of data on a scale far beyond human capabilities. However, many real-world challenges cannot yet be solved solely by AI. Currently, AI lacks the deep conceptual and emotional understanding humans possess about objects and experiences~\cite{de2023ai}. AI cannot interpret human language's nuances, including contextual, symbolic, and cultural meanings~\cite{nath2020problem, mitchell2021abstraction}. AI cannot be an ethical decision-maker because it lacks the human attributes of intentionality, care, and responsibility~\cite{de2023ai}. Therefore, combining human insight with AI's analytical power is crucial for addressing complex real-world challenges, leveraging the strengths to compensate for each other's weaknesses.

\subsection*{Human-AI hybrid collective intelligence}

Even though the field of collective intelligence initially focused solely on groups of people, in recent years, it has gradually expanded to include AI as group members in a new framework referred to as ``hybrid intelligence" \cite{dellermann2019hybrid}. Proponents of the hybrid intelligence perspective stress that humans and AI can connect in ways that allow them to collectively act more intelligently and achieve unreachable goals by any individual entities alone~\cite{akata2020research}. Researchers investigating this hybrid collective intelligence explore ``how people and computers can be connected so that collectively they act more intelligently than any person, group, or computer has ever done before"~\cite{malone2018human}. 

Acknowledging the complementary capabilities of humans and AI as discussed in the previous section, researchers identify the need for developing socio-technological ensembles of humans and intelligent machines that possess the ability to collectively achieve superior results and continuously improve by learning from each other~\cite{dellermann2021future}. Previous research provides evidence that teaming humans with AI has the potential to achieve complementary team performance (CTP), a level of performance that AI or humans cannot reach in isolation~\cite{mcneese2021my}. For instance, a study demonstrates that humans can use contextual information to adjust the AI’s decision~\cite{hemmer2022effect}. Research on mixed teams composed of humans and AI shows hybrid teams could achieve higher performance metrics, such as team situational awareness and score, than all-human teams~\cite{mcneese2021my}. Finally, recent advancements in developing Large Language Models (LLMs) have also been proposed to reshape collective intelligence~\cite{nhb}.

\subsection*{Empirical evidence, theoretical gaps, and a new framework}

Technological advances have constantly disrupted how we produce, exchange, collect, and analyze information and consequently make decisions, individually or collectively. Automation through smart devices based on machine learning, knowledge graph-based machine reasoning, and walking and talking household devices leveraging natural language processing, have already reshaped our personal, professional, and social lives. Collective intelligence cannot be an exception. Inevitably, our collective decision-making processes have been, and will be, disrupted by AI.

For example, machine learning and automation can increase the efficiency and scalability of collective intelligence in Citizen Science projects, where citizens volunteer to help scientists tag and classify their large-scale datasets~\cite{jackson2018folksonomies}. Members of a crowd with different interests and areas of competence can be matched to other tasks by recommendation systems~\cite{yuen2011task}. Generative algorithms can extrapolate human solutions and generate new ideas~\cite{scirea2015moody}. Clustering algorithms can reduce a complex task's solution space as humans explore possible solutions~\cite{hocking2018automatic}. Machines can unify similar solutions to mitigate statistical noise~\cite{zhang2015multi}. Matching algorithms can match individuals and build efficient groups~\cite{karger2011iterative}. 

In this review of AI-enhanced collective intelligence, we adopted a narrative review methodology guided by a complexity theory and network science framework. We began with a conceptual narrative and systematically identified relevant literature for each section. Therefore, we do not claim we have exhausted the vast and fast-growing literature. However, our approach integrates diverse theoretical perspectives and empirical studies, using these theories to explain collective intelligence and its enhancement by AI. This approach bridges interdisciplinary insights, offering a holistic understanding of AI-CI.

This review focuses on integrating AI to bolster the collective intelligence of human groups, addressing a notable theoretical gap in understanding performance enhancements, particularly in hybrid human-AI configurations. With AI integration becoming pervasive across various sectors, exploring how this collaboration can unlock optimal capabilities is imperative. This exploration is essential for boosting productivity, fostering innovation, and ensuring that AI complements and enhances human skills rather than replacing them outright. The industry's pivotal role in driving current AI-CI research underscores the absence of a comprehensive theoretical framework~\cite{berditchevskaia2022descriptive}.

In subsequent sections, we introduce a framework to deepen our comprehension of AI-enhanced collective intelligence systems. We then elaborate on the applications and discuss the implications of such fusion. Finally, we conclude by addressing existing challenges in designing AI-CI systems and speculate on the field's future trajectory.

\section*{AI-enhanced collective intelligence framework}

\subsection*{Multilayer representation of the collective intelligence system}  %human-AI complex

In advancing the field of human-AI collective intelligence, prior research shows the necessity of formulating theories encompassing collective intelligence, combining both humans and AI \cite{eide2016human}. Developing such theoretical frameworks requires an in-depth comprehension and interpretation of human-AI systems, characterized by their high complexity and interrelated processes. Network science offers tools that enable us to understand the complexity of social systems~\cite{pedreschi2023social} where a multitude of interacting elements give rise to the collective behavior of the whole system~\cite{bianconi2023complex}. 
To enhance our understanding of collective intelligence in the human-AI system and explore how ``the whole is more than the sum of its parts"~\cite{aristotle1933metaphysics, anderson1972more}, we integrate complexity science and network science approaches and propose a multilayer representation of the complex system involving human and AI agents.

%\begin{figure}[htbp]
%    \begin{center}
%      \includegraphics[scale=0.9]{new_multiplex.pdf} %0.08
%    \end{center}
%    \caption{A multilayer representation to untangle the processes in the complex system of human and AI agents, with three layers influencing each other: the cognition, physical, and information layers. External factors and the changing environment can influence the emergent collective intelligence of the whole system.} 
%    \label{fig:fig1}
%\end{figure}

Complex system thinking has been used to understand diverse biological, physical, and social domains~\cite{newman2018networks, peters2014application, rosas2017systems}. Inspired by this approach, a real-world collective intelligence system can be mapped into a multilayer network with three interconnected layers: cognition, physical, and information.  In this network, nodes represent interacting agents, and links represent relationships between them. The construction of such networks and the meaning of the links are context-dependent. A node can also represent a group of agents instead of an individual agent in situations where the group behaves or is treated as a single entity within the network. Figure 1 illustrates a multiplex network~\cite{sola2013eigenvector}, a special type of multilayer network where nodes remain the same but links differ across layers. %\ref{fig:fig1}

%\begin{table}[h]
 %   \centering
    % Use 'p{width}' to specify column width and left alignment
  %  \begin{tabular}{c p{4cm} p{9cm}}  % Adjust 10cm to fit your document layout
   %     \hline
    %    \textbf{Layer} & \textbf{Nodes and Links}   & \textbf{Descriptions} \\
     %   \hline
      %  Cognition  &  Farmer A - Farmer B &  
       % \makecell[{{p{9cm}}}]{\raggedright Farmer A reads and interprets Farmer B’s post on pest control. \\ 
       % Farmer A evaluates the advice's effectiveness. \\
       % Farmer A decides on the best method to adopt.}\\
       % \hline
       % Information  &  Farmer A - Farmer C   Farmer C - Farmer D  Farmer D - Farmer A & 
       % \makecell[{{p{9 cm}}}]{\raggedright
       %  Farmer A sends a message to Farmer C about purchasing pest control equipment.\\ Farmer C is in hospital and forwards the message to Farmer D.\\ Farmer D agrees to meet Farmer A for the trade. \\ After the trade, Farmer D asks Farmer A for follow-up messages or reviews.}\\
       % \hline
       % Physical  &  Farmer A - Farmer D &  
       % Farmer A meets with Farmer D to buy the equipment and completes the purchase.\\
       % \hline
    %\end{tabular}
    %\caption{A simplified illustration of a real-world collective intelligence system's layers, nodes, and links, using WeFarm~\cite{WeFarm} as an example. Note that the links can occur at different timestamps, and interactions in one layer can trigger interactions in others. In speculative future scenarios, Farmer D could be replaced by an AI-powered robot that meets with Farmer A to complete the purchase.}
    %\label{tab:wefarm}
%\end{table}

The cognition layer, which contains mental processes, is only indirectly perceivable. Intelligent processes are involved during problem-solving, such as sense-making, remembrance, creation, coordination, and decision-making, often happening in the cognition layer. The links are hard to determine, but these processes exist and are fundamental to the emergence of collective intelligence in the whole system. In opposition to this indirectly perceivable layer is the physical layer, where humans and AI have tangible physical interactions. 
Information interaction in the information layer refers to exchanging information between agents through various communication channels. Besides intra-layer interactions, interactions in one layer can lead to information transfer and trigger interactions in other layers, represented as inter-layer links in the network~\cite{boccaletti2014structure}. This interdependence and cross-layer influence are essential features of multilayer networks. Moreover, Even though our current framework revolves around pairwise interactions between nodes, recent advancements in the study of higher-order networks, going beyond pairwise interactions~\cite{battiston2020networks}, may enrich our understanding of the collective emergence of intelligence.

%Table \ref{tab:wefarm} illustrates the multilayer structure of a real-world collective intelligence system, exemplified by WeFarm~\cite{WeFarm}, a peer-to-peer platform enabling small-scale farmers to share advice and information via SMS and online services to improve farming practices and productivity.

%In this and similar AI-CI applications (see Section Applications and implications), AI functions mainly as a technical tool, with humans as the primary nodes. While AI can physically interact with humans, such as AI-powered robots in warehouses, current technology mostly supports AI in a supportive role rather than as a fully interactive agent. 

In addition to internal processes, it is essential to acknowledge the influence of the environment on the system~\cite{janssens2022collective}. The human-AI complex system functions in a potentially dynamic environment. Previous research suggests that collective intelligent behavior depends on the environment~\cite{janssens2022collective}, as certain behaviors or strategies that work well in one setting may not be as effective or relevant in another~\cite{kammer2014adaptive}.

\begin{table}[ht]
    \centering
    % Use 'p{width}' to specify column width and left alignment
    \begin{tabular}{c p{9cm}}  % Adjust 10cm to fit your document layout
        \hline
        \textbf{Components} & \textbf{Features} \\
        \hline
        Layers & \makecell[{{p{9cm}}}]{\raggedright Cognition layer \\ 
                Information layer \\
                Physical layer }\\
        \hline
        Nodes & \makecell[{{p{9cm}}}]{\raggedright \textbf{Human}: surface-level, deep-level diversity \\ 
                \textbf{AI agents}: diversity in functionality and anthropomorphism} \\
        \hline
        Links & \makecell[{{p{9cm}}}]{\raggedright Links can have directions. \\
                \textbf{Intra-layer links}: interactions within layers \\
                \textbf{Inter-layer links}: interactions between layers} \\
        \hline
        Structure & \makecell[{{p{9cm}}}]{\raggedright Size, centrality, density, hierarchy, community ...} \\
        \hline
        Dynamics & \makecell[{{p{9cm}}}]{\raggedright Communication patterns \\ 
                Cognitive processes \\ etc} \\
        \hline
    \end{tabular}
    \caption{Summary and breakdown of the multilayer network of the human-AI complex system.}
    \label{tab:tab1}
\end{table}

Human-AI hybrid systems can be viewed as complex adaptive systems, continually evolving and adapting through interactions within dynamic environments~\cite{goldstein2011emergence}. 
The concept of emergence in complex systems can be used to describe the phenomenon where new properties emerge at the collective level, which is not present at the level of individual components~\cite{artime2022origin, goldstein2011emergence, meehl1956concept, girvin2019modern}. 
Here, the collective intelligence of the whole system can be seen as
%is such 
an emergent property superior to the micro-level individual intelligence. 
This property, which includes outcomes, abilities, characteristics, and behaviors, is aligned with and encompasses the major existing definitions of collective intelligence.
It emerges through complex nonlinear relationships between the agents and is likely a result of bottom-up and top-down processes~\cite{woolley2015collective}. The former encompasses the aggregation of group-member characteristics that foster collaboration. The latter includes structures, norms, and routines that govern collective behavior, thereby influencing the efficacy of coordination and collaboration~\cite{woolley2015collective}. 
The modeling of the emergence of collective intelligence has leveraged analogies from a diverse array of fields~\cite{galesic2023beyond}, ranging from statistical physics~\cite{de2017criticality} to neuroscience~\cite{daniels2017dual}, both within and beyond.

Employing a multilayer representation of this complex system in a changing environment can facilitate a deeper understanding of the interactions and relationships between humans and AI agents, untangling the intricate processes involved in emerging and maintaining collective intelligence.
Given the interdisciplinary nature of this approach and the variety of terminologies used, key terms and main concepts are listed in supplemental information Table S1 for more clarity.
Table \ref{tab:tab1} provides a summary and breakdown of the components and features within the proposed multilayer network.
The immediate benefit of this framework is that we can learn from the vast literature on multilayer networks and the studies on their robustness\cite{kumar2020robustness}, adaptivity\cite{aleta2019multilayer}, scalability\cite{interdonato2020multilayer}, resilience\cite{demeester1999resilience}, and interoperability\cite{fortino2018towards}. We will discuss this further in the following sections. 

\subsection*{The goal and the task} 

In the previous section, we viewed the hybrid human-AI groups as complex systems that often function in dynamic and complex task environments. Addressing challenges within such systems necessitates a clear understanding of the specific tasks' goals and nature. The system's goal might diverge from individual objectives, highlighting the need for strategic coordination to balance collective aims with personal pursuits. 

Task types can vary significantly, from generative activities such as brainstorming, which requires creative and divergent thinking, to analytical tasks such as solving Sudoku, which demands logical reasoning and pattern recognition~\cite{delahaye2006science}. The diversity of the tasks requires different abilities, including memory utilization, creative imagination, sense-making, and critical analysis. In the following subsections, we focus on the details of group-member diversity's influence, the structure and dynamics of interaction networks, and AI's role in enhancing group performance. 

\subsection*{Group diversity} 

The relationship between group diversity and collective intelligence is highly complex due to diversity's complexity and the range of its effects under different conditions. 
A group can exhibit surface-level %(demographic) and deep-level %(psychological) 
diversities~\cite{phillips2006surface}. The surface-level diversity comes from readily observable social categories, including gender, age, and ethnicity~\cite{wegge2009impact}. 
The deep-level diversity refers to differences in psychological characteristics~\cite{harrison2002time}, including personality, cognitive thinking styles, and values. 

Diversity benefits teams by enhancing creativity, improving decision-making, and expanding access to a broader talent pool. ~\cite{bagshaw2004diversity, van2003joint}.  However, diversity can also have adverse effects, such as emotional conflict~\cite{pelled1999exploring,yasseri2012dynamics}, stress~\cite{defrank1998stress}, poor work relationships, and poor overall performance~\cite{van2004work}.
There is a lack of consensus on which aspects of group diversity are likely to result in positive outcomes and which aspects could produce negative outcomes~\cite{van2004work, jansen2021diverse}. 
The effects of diversity on group performance depend on the complexity of the task and various moderators, such as the density of the interaction network~\cite{baumann2024network}. A review underscores the need for in-depth research into surface and deep-level diversity within hybrid teams~\cite{jansen2021diverse}, exploring their impact across various outcomes and contexts over time.

\subsubsection*{\textbf{Surface-level diversity}}

\paragraph{\textbf{Gender}}
 
The evidence concerning the effect of gender diversity on team performance is equivocal and contingent upon various contextual factors~\cite{bear2011role}. 
Past research found that gender-diverse teams outperform gender-homogeneous teams when perceived time pressure is low~\cite{kearney2022gender}.
One study showed gender diversity was negatively associated with performance, but only in large groups~\cite{wegge2009impact}. 
Previous evidence strongly suggests that team collaboration is greatly improved by the presence of females in the group~\cite{bear2011role, woolley2010evidence}. 
Gender-diverse teams produce more novel and higher-impact scientific ideas~\cite{yang2022gender}. 
Other studies show that gender composition seems to matter for team performance. However, when controlling for the individual abilities of team members, the relation between gender composition and team performance vanishes~\cite{stadter2022differences}.
One study points out that the ``romance" of working together can benefit group performance~\cite{zhang2012romance}. Gender-diverse groups perform better than homogeneous groups by decreasing relationship and task conflict~\cite{zhang2012romance}.

\paragraph{\textbf{Age}} 

Age diversity may enhance group performance, but the benefits may depend on the context.
Several studies have found that team age diversity positively correlated with team performance when completing complex tasks~\cite{wegge2012makes}. 
Past research points out the positive effects were only seen under conditions of positive team climate and low age discrimination~\cite{wegge2012makes}. 
Another study shows teams with more age diversity reported more age discrimination~\cite{kunze2013organizational}, associated with lower commitment and worse performance.

\paragraph{\textbf{Ethnicity}} 

There are inconsistent results in the case of ethnic diversity. 
One study~\cite{mcleod1996ethnic} found racially diverse groups produced significantly more feasible and effective ideas than homogeneous groups. 
Another study found racial similarity in groups associated with higher self-rated productivity and commitment~\cite{riordan1997demographic}. 
Under the right conditions, teams may benefit from diversity in ethnicity and nationality~\cite{nielsen2017gender}. 

\paragraph{\textbf{AI surface-level diversity}}

AI exhibits surface-level diversity through different anthropomorphic features in robots, such as voice, avatars, and human-like characteristics. Research has indicated that the gendering of AI can affect perceptions and trust levels in users. For instance, studies have found that users may exhibit more trust in AI personal assistants whose voice gender matches their own~\cite{lee2021social}.
Assigning a female gender to AI bots enhances their perceived humanness and acceptance~\cite{borau2021most}. 
Previous studies have indicated a perception bias in AI agents based on gender: male AI agents tend to be viewed as more competent, whereas female AI agents are often perceived as warmer~\cite{ahn2022effect}. 
The perceived gender of the machine can make the social dynamics in hybrid teams even more complicated. However, these gendered characteristics in AI also raise concerns about reinforcing stereotypes and biases~\cite{manasi2022mirroring}. The influence of AI’s gendered traits on human interactions is an ongoing area of research, with implications for how AI is designed and utilized.
Despite substantial research on AI gender, there is a notable gap in understanding the influence of other characteristics, such as AI age and ethnicity, on human interactions, highlighting the need for further exploration in this area.

\subsubsection*{\textbf{Deep-level diversity}}

\paragraph{\textbf{Personality}} 

Some work proposes that investigations into the relationship between personality and work-related behaviors should expand beyond the linearity assumptions~\cite{curcseu2019personality}, showing that extraversion, agreeableness, and conscientiousness have inverted U-shaped relationships with peer-rated contributions to teamwork~\cite{curcseu2019personality}. Similar work demonstrates a curvilinear relationship between a team's average proactive personality and performance~\cite{zhang2021too}. This body of work also identifies the moderating effects of personality dispersion, team potency, and cohesion on this relationship~\cite{zhang2021too}. 
Previous research has found that a team's collective openness to experience and emotional stability moderates how task conflict affects team performance~\cite{bradley2013ready}. Specifically, in teams with high openness or emotional stability, task conflict positively influences performance~\cite{bradley2013ready}.

\paragraph{\textbf{Cognitive style}} 
 
Team cognitive style diversity refers to the variation in team members' ways of encoding, organizing, and processing information~\cite{aggarwal2019impact}. Research indicates its significance in fostering innovation within teams~\cite{aggarwal2019team}. A certain level of cognitive diversity contributes to collective intelligence by providing diverse cognitive inputs and viewpoints necessary for task work~\cite{aggarwal2019team}. However, excessive diversity may lead to high coordination costs~\cite{straub2023cost}, as team members with differing perspectives struggle to understand each other~\cite{cronin2007representational, mello2015cognitive}. Studies suggest that the relationship between cognitive style diversity and collective intelligence is curvilinear, forming an inverted U-shape~\cite{aggarwal2019impact}. Additionally, the beneficial effects of cognitive diversity might be mediated through factors such as task reflexivity and relationship conflict~\cite{chen2019cognitive}.

\paragraph{\textbf{Value judgment}} 

Values are internalized beliefs that can guide behavior and enhance motivation~\cite{cennamo2008generational}. 
Past research found higher diversity in values was associated with more conflict~\cite{liang2012impact} in relationships, tasks, and processes.  
Overall, existing research consistently indicates that value congruence within teams positively influences performance, satisfaction, and conflict~\cite{jansen2021diverse} and moderates the effects of informational diversity~\cite{jehn1999differences}.

\paragraph{\textbf{AI deep-level diversity}}

While AI systems can display various behaviors or responses based on their programming and training, this diversity is a product of different algorithms, models, and data inputs. Unlike humans, AI does not have personal experiences, innate personality traits, cognitive thinking, or beliefs contributing to deep-level diversity. Although AI lacks inherent deep-level traits, AI might influence deep-level diversity in human teams, primarily by shaping social interactions and decision-making processes. The optimal composition of teams remains uncertain~\cite{bell2007deep}.

\subsection*{Interactions}  

Interaction distinguishes a group from a mere collection of individuals~\cite{janssens2022collective}: one person's behavior forms the basis for the responses of others~\cite{driskell1992collective}. Therefore, the interaction between two members establishes a link in the interaction network. 
In terms of the group’s interaction network structure, past research suggests that the small-worldness of a collaboration network (small network diameter and high clustering) improves performance up to a certain point, after which the effect will reverse~\cite{uzzi2005collaboration}. 
An ``inverted U" relationship is found between connectedness and performance~\cite{lazer2007network}, suggesting an optimum number of connections for any given size of the collaboration network. 
A modular structure (a large number of connections within small groups that are loosely connected) can increase efficiency~\cite{navajas2018aggregated}.  
A study on forbidden triads among jazz groups shows that heavily sided open triangles are associated with lower success (measured by the number of group releases)~\cite{vedres2017forbidden}. 

AI can contribute to human groups in various ways by augmenting existing human skills or complementing capabilities that humans lack. 
AI has attributes beyond humans, such as more extended memory, higher computational speed, and a more vital ability to work with large, diverse datasets. 
AI's augmentation of human cognition enables teams to navigate complex situations effectively~\cite{woolley2023using}. 
AI may also help reduce the harm of implicit bias in humans~\cite{lin2021engineering} and shape better decisions. Taking advantage of the differences between humans and AI can contribute to the performance of hybrid human-AI groups.
In the following sections, we discuss the rules and incentives that govern the interaction processes and the structure and dynamics of the group's interaction network.

\subsubsection*{\textbf{Rules and incentives}}  

Decision rules function as prescribed norms that direct interactions and play a crucial role in shaping the communication and information integration within a group~\cite{meslec2014none}. 
Previous research has focused on disentangling decision rules guiding the team’s interaction, ultimately fostering synergy. 
Some literature emphasizes that the decision rules themselves are intelligent~\cite{wolf2015collective}.
In addition to the rules guiding behavior, the group members interact with each other under specific incentive schemes. 
The intrinsic and extrinsic incentives are the driving forces~\cite{augmentedCollectiveIntelligence2024} for the agents to behave in a direction such that the social network becomes dynamic and moves towards a collective goal. 
Some research proposes that an incentive scheme rewarding accurate predictions by a minority can foster optimal diversity and collective predictive accuracy~\cite{mann2017optimal}.

AI can augment human cognition~\cite{regens2019augmenting} to help teams adapt to complexity. 
While AI can simulate certain human-like incentives through programming and learning algorithms~\cite{park2023generative}, fully capturing the depth and nuance of human behavior, such as irrationality or altruism, is challenging. A recent study applying a Turing test to compare AI chatbots and humans found that AI chatbots exhibited more altruistic and cooperative behavior than their human counterparts in behavioral games~\cite{mei2024turing}. Machines can be programmed to mimic human behaviors, but whether they genuinely ``exhibit" them like humans is debatable. In collective intelligence, integrating such human-like behaviors in AI may enhance collaboration and decision-making within diverse teams. However, the extent to which AI can be truly self-aware and authentically replicate complex human traits remains a subject of ongoing research and philosophical debate.

%\newpage
\subsubsection*{\textbf{Group structure and dynamics}} 

\paragraph{\textbf{Group size}} 

Much research on collective intelligence has been dedicated to finding a group's optimum size and structure. 
Several studies have reported that performance improves with group size through enhanced diversity~\cite{malone2022handbook}, whereas others suggest the opposite: small groups are more efficient~\cite{galesic2018smaller}. Another study suggests that large teams develop and small teams innovate~\cite{wu2019large}.
Larger groups can better utilize diverse perspectives and knowledge to solve complex problems but also tend to experience more coordination problems and communication difficulties~\cite{woolley2010evidence}. 
The consideration of optimal group size can be related to the dynamics of the group, whether it is static, growing, or diminishing.

In human-AI hybrid teams, group size becomes nuanced with AI integration, especially when AI entities such as LLMs are involved. The countability of AI agents, particularly if multiple agents are derived from a single LLM, raises questions about their distinctiveness and individuality in a team context. Unlike humans, AI agents can simultaneously process multiple tasks, blurring the traditional boundaries of group size. Applying past research on human group dynamics to these hybrid systems requires rethinking the notions of individual contribution, team cohesion, and communication. It is important to consider how AI's unique capabilities and scale impact these dynamics and whether multiple AI agents from a single model represent distinct entities or a collective intelligence resource.

\paragraph{\textbf{Structure}}

Research shows that network structure, specifically hierarchy and link intensity, can affect information distribution~\cite{tang2023group}, influencing group performance.
Another study observed that the presence of structural holes~\cite{burt2018structural} in leaders' networks and the adoption of core-periphery and hierarchical structures in groups correlated negatively with their performance~\cite{cummings2003structural}.
Incorporating the temporal dimension, adaptive social networks can significantly impact collective intelligence by allowing the network's structure to evolve based on feedback from its members~\cite{almaatouq2020adaptive}. 

Team size moderates the relationship between network structure and performance~\cite{yuan2022leader}. 
Studies have found central members can coordinate with other team members more easily in smaller teams~\cite{balkundi2006ties}, whereas in larger teams, communication challenges may arise~\cite{hollenbeck2002structural}. 
A recent study shows a more central connected leader in advice-giving networks has a more positive impact on the performance of larger teams, but this effect is reversed in smaller teams~\cite{yuan2022leader}.

Determining the optimal size, structure, and human-AI ratio for hybrid teams requires further study, particularly concerning task-specific requirements. Moreover, incorporating AI into human teams can reshape group hierarchies, norms, and rules, affecting collective intelligence.

\paragraph{\textbf{Communication patterns}} 

Teams often outperform individuals primarily due to explicit communication and feedback within the team~\cite{larson2013search}. 
Understanding the communication network in crowds is crucial for effectively designing and managing crowdsourcing tasks~\cite{yin2016communication}.
Various studies highlight that centralized and decentralized communication patterns can effectively promote team performance, contingent on the nature of the task and the team's composition~\cite{cohen1962effects, rosen2008cooperation}. 
In particular, teams handling complex tasks tend to be more productive with decentralized communication networks~\cite{shaw1964communication}. 
Conversely, the centralization of communication around socially dominant or less reflective individuals can negatively impact the utilization of expertise and overall team performance~\cite{sherf2018centralization}. 
Communication patterns, such as equal participation and turn-taking of team members, can positively affect collective intelligence~\cite{woolley2010evidence}.

The integration of AI in human groups can influence the conversation dynamics in different ways, depending on the role of AI. AI chat interventions can improve online political conversations at scale~\cite{argyle2023leveraging}, reducing the chance of conflicts. 
While AI machines might struggle with capturing the subtle and ineffable social expressions that make up the dynamics of human groups, this can be advantageous. In certain contexts, AI can be social catalysts~\cite{rahwan2020intelligent} to promote communication where human capacity is limited. Striking the right balance in AI's role is key to maintaining the natural dynamics of human interaction.

\paragraph{\textbf{Cognitive processes}} %Cognition
 
Cognitive processes are foundations for the emergence of collective intelligence~\cite{woolley2024understanding}, encompassing knowledge acquisition, information processing, problem-solving, decision-making, language, perception, memory, attention, and reasoning. 
Human cognition processes evolve and interact with each other during various tasks and experiences. 
Humans can represent the mental states of others, including desires, beliefs, and intentions~\cite{rabinowitz2018machine}, known as the Theory of Mind (ToM). 
Past research, inspired by ToM, showed that team members' ability to assess others' mental state is positively associated with team performance both in face-to-face settings~\cite{woolley2010evidence} and online~\cite{engel2014reading}. 
Regarding collective understanding within a group, the Shared Mental Models (SMM) theory is helpful for understanding, predicting, and improving performance in human teams~\cite{andrews2023role}.

Applying theories such as SMM to human-AI teams is a natural progression in research. These models can enhance team performance by ensuring that human and AI members have a similar understanding of tasks and each other~\cite{andrews2023role}, facilitating better prediction of needs and behaviors. Developing ToM in AI also improves coordination and adaptability in human-AI teams~\cite{baker2011bayesian}. This involves AI's capability to anticipate and respond to new information and behaviors. Moreover, recent studies indicate that incorporating ``hot" cognition, influenced by emotional states in AI, can further improve human-machine interactions~\cite{cuzzolin2020knowing}. Understanding the transactive systems~\cite{woolley2023using} of memory, attention, and reasoning within these teams is key to grasping the members’ knowledge and task preferences.

\subsection*{AI modes of contribution}

AI's role in hybrid groups varies based on its autonomous agency and functionality. AI may serve as a technical tool for human assistance or as an active agent that interacts with and influences humans. Functionally, AI's role can range from an assistant to a teammate, coach, or manager. Furthermore, the degree of AI's anthropomorphism, from non-physical voice-only interfaces to avatars and physically embodied robots, also significantly affects its role in the group.
Table \ref{tab:ai_roles} outlines the key roles of AI in human-AI contexts, accompanied by descriptions.

\begin{table}[ht]
    \centering
    \begin{tabular}{c p{7.5cm} p{5cm}}
        \hline
        \textbf{Role} & \textbf{Description} & \textbf{Examples} \\
        \hline
        \makecell[c]{Assistant} & Complements or augments human abilities by performing tasks such as language translation, administrative duties, and smart home device control. & Google Translate; Alexa or Siri managing smart home devices. \\
        \hline
        \makecell[c]{Teammate} & Collaborates with humans, offering complementary skills and enhancing team performance in various fields, such as healthcare and creative industries. & AI collaborating with radiologists in image diagnosis; Google’s Magenta collaborating with artists. \\
        \hline
        \makecell[c]{Coach} & Provides guidance, feedback, and strategic oversight, helping individuals and teams improve their skills and coordination. & AI coaching in team sports, providing strategic advice; AI mentoring employees in skill development. \\
        \hline
        \makecell[c]{Manager} & Assists in decision-making processes, reduce biases, promotes diversity, and optimizes task allocation and team dynamics. & AI in hiring and promotion decisions; AI optimizing task allocation in project management. \\
        \hline
        \makecell[c]{Embodied Partner} & Integrates AI with robotics, enabling physical interactions and augmenting human capabilities in tasks requiring a physical presence. & Robotic arms in factories; autonomous vehicles aiding in logistics and transportation. \\
        \hline
    \end{tabular}
    \caption{Key roles of AI in hybrid groups}
    \label{tab:ai_roles}
\end{table}

%\subsubsection{\textbf{AI as an assistant}}
Assistant-type AI systems generally function as technical tools with limited autonomy, designed to complement or augment human abilities and enhance efficiency in task performance.
For example, language translation AI can assist in translating text or spoken language, facilitating communication across language barriers. 
%Generative AI can inspire human creators and provide them with new ideas \cite{doshi2023generative}.  
LLMs in education can assist students' learning by adapting to different roles based on the prompts provided by the students~\cite{mollick2023assigning}.
Administrative AI assistants can help schedule meetings, manage emails, and organize tasks. 
Smart home AI assistants such as Alexa and Siri can help manage smart home devices, play music, provide weather updates, answer questions, and assist with daily routines.
AI assistants are technical tools that help humans coordinate tasks efficiently, optimize their decision-making, and personalize their experiences.

%\subsubsection{\textbf{AI as a teammate}} 

AI can function as teammates, working alongside humans with complementary abilities. AI teammates are already used in real-world settings; most employees using AI already see it as a coworker.
In healthcare, radiologists and AI work together to diagnose from the pictures of the patients. In creative industries, AI collaborators in music, art, and literature, such as Google's Magenta project, collaborate with musicians and artists to create new compositions or artworks using AI algorithms. 
Furthermore, AI robots can serve as teammates in various workplaces, a topic we will discuss in more detail later.
% in the AI embodiment section.
   
%\subsubsection{\textbf{AI as a coach}} 

As a coach, AI can provide guidance, support, and personalized feedback to individuals seeking to improve their skills and achieve their goals. 
In team settings, one perspective highlights the AI coach's ability to provide a comprehensive, global view of the team's environment, guiding players who have only partial views~\cite{liu2021coach}. This approach involves the AI coach strategizing and coordinating, distributing tailored strategies to each team member based on their unique perspectives and roles. Another dimension of AI coaching focuses on assessing and improving teamwork by closely monitoring team members during collaborative tasks~\cite{seo2021towards}. Here, the AI coach actively intervenes with timely suggestions based on the team members' inferred mental model misalignment~\cite{seo2021towards}.

%\subsubsection{\textbf{AI as a manager}} 

AI has been used in working environments with supervision or managerial roles, such as making decisions of hiring, promotion, and reassigning tasks~\cite{Kiron2023Workforce}. 
An AI manager may mitigate human biases~\cite{hofeditz2022applying} in hiring and promotions, contributing to a more diverse workforce, which is a key aspect of collective intelligence. Additionally, by analyzing individual strengths and weaknesses and efficiently allocating tasks, an AI manager can optimize team dynamics and workflows, improving overall group performance.

%\subsubsection{\textbf{AI embodiment}} 

With the advancement of AI in areas such as natural language processing, object recognition, and creative idea generation, there's a trend to integrate AI with robotics, creating physical entities in the real world. 
Embodied AI~\cite{pfeifer2004embodied} refers to intelligent agents interacting with their environment through a physical body. 
These AI-enabled machines, embodied in robotic arms or autonomous vehicles, can augment the capabilities of human workers. 
They are equipped with advanced sensors, motors, and actuators, enabling recognition of people and objects and safe collaboration with humans in diverse settings, including factories, warehouses, and laboratories. 
Furthermore, advanced AI-robotics integration allows robots to recognize and respond to human speech and gestures, impacting conversational dynamics within mixed human-robot environments.
Research has shown that a vulnerable robot's social behavior positively shapes the conversation dynamics among human participants in a human-robot group~\cite{traeger2020vulnerable}. In scenarios where conflicts emerge within a human group or in groups of strangers, a humorous robot can potentially serve as an ice breaker to intervene and resolve the awkward situation~\cite{press2023humorous}, demonstrating the potential of robots to influence the nature of human-human interactions notably.

\subsection*{Human-AI collective intelligence emerging factors}

\subsubsection*{\textbf{Perception and reaction}}

In addition to investigating how humans perceive individual AI, previous research suggests the necessity of additional work to explore human perceptions of AI in collaborative team environments~\cite{kocielnik2019will}. 
For human-AI teams to succeed, human team members must be receptive to their new AI counterparts~\cite{harris2023social}. 
People's pre-existing attitudes toward AI were found to be significantly related to their willingness to be involved in human-AI teams~\cite{zhang2021ideal}. 
Research finds that two aspects of social perception, warmth, and competence, are critical predictors of human receptivity to a new AI teammate. Psychological acceptance is positively related to perceived human-AI team viability~\cite{harris2023social}. 
Regarding human perception of AI behaviors, one study finds that people cooperate less with benevolent AI agents than with benevolent humans~\cite{karpus2021algorithm}. 

Humans tend to be more accepting of AI when it makes mistakes. Some studies find that users have a more favorable impression of the imperfect robot than the perfect robot when the robot behaves adequately after making mistakes~\cite{yasuda2013psychological}. 
When errors occur in human-machine shared-control vehicles, the blame assigned to the machine is reduced if both drivers make mistakes~\cite{awad2020drivers}.
Research has found that errors occasionally performed by a humanoid robot can increase its perceived human likeness and likability~\cite{salem2013err}. 
A recent study shows humans accept their AI teammate's decision less often when they are deceived about the identity of the AI as another human~\cite{zhang2023trust}. 
The nature of the task requested by the robot, e.g., whether its effects are revocable as opposed to irrevocable, significantly impacts participants’ willingness to follow its instructions~\cite{salem2015would}.  
When taking AI's advice, humans must know the AI system’s error boundary and decide when to accept or override AI’s recommendation~\cite{bansal2019beyond}. 

\subsubsection*{\textbf{Trust}} 

In AI-assisted decision-making, where the individual strengths of humans and AI optimize the joint decision outcome, the key to success is appropriately calibrating human trust~\cite{zhang2020effect}. 
Trust between humans and AI  plays a pivotal role in realizing collective intelligence~\cite{gupta2023fostering}. 
Previous research highlights the necessity of understanding factors that enable or hinder the formation, maintenance, and repair of trust in human-AI collaborations~\cite{gupta2023fostering}. Key factors influencing this trust include the perceived competence, benevolence, and integrity~\cite{mcallister1995affect} of the AI systems, mirroring the trust dynamics in human relationships.

The level of anthropomorphism of machines can be a predictor of the trust level of the human participants~\cite{natarajan2020effects}. 
Studies indicate that while anthropomorphic features in AI agents initially create positive impressions, this effect is often short-lived~\cite{gupta2023fostering}.
Trust tends to decline, especially when the agent's capabilities are not clearly presented and its performance fails to meet users' expectations~\cite{glikson2020human}.

Trust in a teammate may evolve over time irrespective of team performance~\cite{mcneese2021trust}. 
Knowing when to trust the AI allows the human expert to appropriately apply their knowledge, improving decision outcomes in cases where the model is likely to perform poorly~\cite{zhang2020effect}. 
A previous study also proposes a framework highlighting trust relationships within human-AI teams. It acknowledges the multilevel nature of team trust, which considers individual, dyads, and team levels as a whole~\cite{ulfert2023shaping}. 
Another study points out that the trust level of a team needs to be uniform, as uncooperative members could undermine the team's ability to reach intended goals~\cite{de2021trust}.

\subsubsection*{\textbf{Explainability}} 

In human-AI decision-making scenarios, people's awareness of how AI works and its outcomes is critical in building a relationship with the system. AI explanation can support people in justifying their decisions~\cite{ferreira2021human}. 
A study on content moderation found that when explanations were provided for why the content was taken down, removal decisions made jointly by humans and AI were perceived as more trustworthy and acceptable than the same decisions made by humans alone~\cite{wang2023content}. 
Many researchers have proposed using explainable AI (XAI) to enable humans to rely on AI advice appropriately and reach complementary team performance~\cite{hemmer2022effect}. However, some studies suggest that XAI can be associated with a white-box paradox~\cite{cabitza2023rams}, potentially leading to null or detrimental effects.
 
Previous research finds that AI explanations can increase human acceptance of AI recommendations, regardless of their correctness~\cite{bansal2021does}, leading to an over-reliance, threatening the performance of human-AI decision-making. Whereas another study finds that there are scenarios where AI explanations can reduce over-reliance~\cite{vasconcelos2023explanations} using a cost-benefit framework. 
The impact of AI explanations on decision-making tasks is contingent upon individuals' diverse levels of domain expertise~\cite{wang2021explanations}.
Additionally, some studies advocate for the concept of causability~\cite{holzinger2021toward}, which measures whether and to what extent humans can understand a given machine explanation to develop effective human-AI interfaces, particularly in medical AI.

\section*{Applications and implications} 

The AI-enhanced collective intelligence approach is increasingly applied across various domains, offering innovative solutions to challenges such as community response to climate change, environment, and sustainability challenges~\cite{NestaFutureMindsMachines2023}, etc. 
AI serves as a vital decision support tool for policymakers in detecting misinformation~\cite{kou2022crowd} and real-time crisis management, exemplified during the COVID-19 pandemic. Its application extends to high-stake areas, including medical diagnosis~\cite{lebovitz2022engage} and criminal justice~\cite{rigano2019using}. AI's integration into collective intelligence efforts enables significant amplification of its impact and scalability~\cite{verhulst2018and}.

%\begin{figure}[htbp]
%    \begin{center}
%      \includegraphics[scale=0.55]{AI_CI_cases_2024.pdf} 
%    \end{center}
%    \caption{Distribution of AI-enhanced collective intelligence cases by application area based on dataset curated by Supermind Design~\cite{SupermindDesign2023}.} 
%    \label{fig:fig2}
%\end{figure}

Here, we present our analysis utilizing the Supermind Design database~\cite{database}, which includes over one thousand actual examples of AI-enhanced collective intelligence, with 938 categorized into twelve application areas. Notably, while some cases might not directly involve AI or collective intelligence depending on the exact definitions, the database collectively offers a comprehensive overview. It provides insights into the state-of-the-art in AI-CI applications. 
As shown in Fig. 2, the majority of applications, %\ref{fig:fig2}
approximately twenty percent, are found in the public sector and NGOs, followed by high-tech, media, telecommunications, entertainment, and hospitality. Fewer instances are found in supply chain, real estate, and agriculture. It is important to note that several cases classified under ``Public sector, NGO" are private sector initiatives addressing public issues. In the upcoming sections, we will analyze and highlight specific AI-CI examples across these diverse domains. 
The supplemental information Table S2 provides a summary of the CI and AI aspects of these examples.

\subsection*{Public sector, NGO}
Safecity (safecity.in) %~\cite{Safecity} 
is a platform that crowdsources personal stories of sexual harassment and abuse in public spaces. It collaborates with the AI partner OMDENA
%~\cite{Omdena} 
to build solutions to help women at risk of sexual abuse by predicting places at a high risk of sexual harassment incidents utilizing convolutional neural networks and LSTM (Long-Short-Term-Memory). 

Ushahidi (www.ushahidi.com)
%~\cite{Ushahidi} 
is a non-profit online platform for crowdsourcing information, mapping, and visualizing data to respond to and inform decisions on social issues and crises. 
It aggregates crowdsourced information and reports from individuals across the globe. 
Ushahidi is developing new tools to enhance data management and analysis, such as leveraging machine learning techniques to improve the efficiency of processing crisis reports, thus enabling faster response time~\cite{ushahidi_ai}. 

While NGOs and the public sector might seem like fertile grounds for AI-CI applications, it is crucial to consider the potential high-risk issues that may arise and their impact during the design and deployment of such AI-CI systems. A notable instance of a data-driven tool for social good that diverged from its original accuracy and effect is the infamous case of Google Flu~\cite{lazer2014parable}.

\subsection*{High Tech (software)}
Just as AI-enhanced collective intelligence helps address social issues and crises in the public sector and NGOs, it also drives innovation and efficiency in the high-tech industry.
Bluenove (bluenove.com), a technology and consulting company that pioneers massive collective intelligence, focuses on mobilizing communities on a large scale. Their Assembl platform (bluenove.com/en/offers/assembl)
%~\cite{assembl} 
 facilitates open conversations among participants, employing advanced AI technologies such as natural language recognition, semantic analysis, and emotion analysis to categorize content published on the platform automatically. 

Figure Eight 
%~\cite{figure8} 
(formerly known as CrowdFlower
%~\cite{CrowdFlower}) 
was a human-in-the-loop machine learning and AI company acquired by Appen (appen.com)
%~\cite{Appen2023} 
in 2019. 
It leverages crowdsourced human intelligence to perform tasks such as text transcription or image annotation to train machine learning algorithms, which can be used in various applications.

The recent advancements in high-tech technologies, especially the burgeoning field of LLM software, hold the potential for revolutionary developments. Nevertheless, the high risk associated with errors in automated or AI-generated software~\cite{zhang2023critical} necessitates the inclusion of human intelligence in these projects.

\subsection*{Media, telecommunication, entertainment, hospitality} 
Similar to high tech, more traditional information and communication technologies, such as the media sector, harness AI-enhanced CI for purposes such as investigative journalism, historical research, and combating misinformation.

Civil War Photo Sleuth (civilwarphotosleuth.com)
%~\cite{CivilWarPhotoSleuth2023} 
combines facial recognition technology and community effort to uncover lost identities in photographs from the American Civil War era. It leverages a dedicated community with a keen interest in the US Civil War and focuses on more accurately tagging and identifying individuals in historical photographs.

Bellingcat (bellingcat.com)
%~\cite{bellingcat_info} 
is an independent investigative online journalism community specializing in fact-checking and open-source intelligence. It brings together a collective of researchers, investigators, and citizen journalists to publish investigation findings.
Utilizing various AI tools in investigative work, Bellingcat obtains insights from massive open-source data, such as satellite imagery, photographs, video recordings, and social media posts. 

Facing the emerging challenge of deepfake videos, research indicates that a hybrid system combining human judgment with AI models outperforms either humans or models alone in detecting deepfakes~\cite{groh2022deepfake}, underscoring the value of integrating human perceptual skills with AI technology in addressing the complexities posed by deepfake content.

\subsection*{Energy, natural resources}

Swithing to a more industrial environment, in the energy and natural resources sector, collective intelligence combined with AI can play a  role in addressing environmental challenges through community-driven data collection and analysis. 
Litterati~\cite{litterati} is a mobile app and community platform that combats litter by empowering people to take action. 
The app allows the worldwide community to contribute their observations and findings through photographing and documenting occurrences, creating a crowdsourced database of litter data. 
It also provides a platform for individuals to connect, share, and support each other's initiatives.
The platform uses AI, such as machine learning and computer vision techniques, to automatically recognize different objects and materials and efficiently organize the collected litter data. Similarly, OpenLitterMap (openlittermap.com)
%~\cite{OpenLitterMap2023} 
is another example of making a clearer planet using AI-enhanced collective intelligence. 

eBird~\cite{sullivan2014ebird} is a mobile app and community platform for biodiversity and supports conservation initiatives to protect bird species and their habitats. It harnesses the power of a global community of birdwatchers and citizen scientists to collectively report bird observations and contribute to a shared knowledge base.
eBird uses computer vision techniques to identify bird species from submitted photographs and recordings and employs machine learning predictive models to forecast bird migration patterns and species distributions.  

Tackling environmental challenges is crucial, and large-scale, crowd-based AI-enhanced collective intelligence projects play a significant role in this endeavor. Nevertheless, integrating AI with human efforts might dehumanize the work and demotivate human participants~\cite{trouille2019citizen}. This balance between efficiently leveraging AI and maintaining human engagement is a key consideration in such projects.   

\subsection*{Education and academia}

Education and academia are other areas that benefit from AI-enhanced collective intelligence by fostering collaborative learning environments and supporting large-scale research projects. Kialo Edu (kialo-edu.com)
%~\cite{Kialo} 
is an argument mapping and debate site dedicated to facilitating collaborative and critical thinking. 
Kialo uses natural language processing techniques to analyze arguments, cluster topics, and identify discussion patterns. 
It allows users to participate in structured debates and discussions, promoting constructive engagement among students. This interaction helps them deeply understand and analyze the core aspects of the topics under discussion.

Zooniverse (zooniverse.org)
%~\cite{zooniverse} 
is a platform for people-powered research where volunteers assist professional researchers.
Galaxy Zoo is a prominent project within the Zooniverse platform.
This crowdsourced astronomy project invites people to assist in the morphological classification of large numbers of galaxies. 
It uses AI algorithms to learn patterns and characteristics indicative of different galaxy types by training on large datasets previously classified by citizen scientists.
The combination of efforts from both the crowd and AI enables Galaxy Zoo to harness the power of human expertise and the efficiency of AI to scale up the analysis and understanding of galaxies. 

A study analyzing the contributor demographics of the Zooniverse project revealed uneven geographical distribution and a gender imbalance, with approximately 30\% of the citizen scientists being female~\cite{ibrahim2021gender}. This finding highlights concerns about representativeness and diversity in citizen science initiatives, indicating potential biases in the data and insights from such projects. AI can be used to train human contributors~\cite{zevin2017gravity}, to de-bias collective intelligence, and to evaluate and combine human solutions~\cite{kleinberg2018human}. 

\subsection*{Healthcare}
In a similar vein, the healthcare sector also leverages AI-enhanced collective intelligence, addressing challenges such as diagnostic accuracy and medical data analysis, which are critical for improving patient outcomes and advancing medical research.
The Human Diagnosis Project (humandx.org)
%~\cite{humandx} 
is an open online system of collective medical insights that provides healthcare support to patients.  
Human Dx enables healthcare professionals worldwide to contribute expertise and collaborate on complex diagnostic cases. %~\cite{humandx}. 
The platform facilitates idea exchange and mutual learning to ultimately benefit patient care by utilizing machine learning techniques that automatically learn from classifying patterns in crowdsourced data. 
Human Dx makes the most of limited medical resources by harnessing the collective intelligence of human physicians and AI, enabling more accurate, affordable, and accessible care for those in need. 

CrowdEEG~\cite{crowdEEG} is a collaborative annotation tool for medical time series data. The project combines human and machine intelligence for scalable and accurate human clinical EEG data analysis.
It trains machine learning algorithms using feedback from clinical experts and non-expert crowds to perform feature detection or classification tasks on medical time series data.

In healthcare, the use of AI-CI poses the risk of amplifying existing biases in human-only diagnosis once we combine AI with human judgment~\cite{groh2024deep}. Consequently, errors resulting from such biased decision-making can erode patients' trust not only in the AI systems but also in the medical professionals utilizing them.

\subsection*{Financial services} 
AI-enhanced collective intelligence in financial services optimizes investment strategies and improve market predictions, demonstrating the powerful synergy between human expertise and machine learning.
Numerai (numerai.fund)
%~\cite{numerai} 
is an AI-run quant hedge fund built on crowdsourced machine learning models.
Numerai hosts an innovative data science tournament in which a global network of data scientists develops machine-learning models to predict stock markets. 
Based on the combined knowledge of the participants, Numerai combines and aggregates these crowdsourced predictive models into an ensemble model to derive investment strategies in the financial markets. 

CryptoSwarm AI 
is a forecasting service that provides rigorous insights and intelligence on cryptocurrencies and other Web3 assets. 
A combination of AI technology and real-time human insights powers it. 
CryptoSwarm AI uses Swarm AI
to amplify the collective intelligence of the online communities quickly, enabling networked groups to converge on optimized solutions in real time. 

In financial markets, research generally indicates positive impacts on market efficiency due to the superior information processing and optimization capabilities of machines, coupled with human anticipation of these effects. However, this improvement is sometimes disrupted by rapid price spikes and crashes, resulting from the herding behavior of machines and the challenges humans face in intervening at extremely fast timescales~\cite{tsvetkova2024human}.

\subsection*{Supply chain, real estate}
The supply chain and real estate sectors also benefit from AI-enhanced collective intelligence, improving efficiency, transparency, and decision-making through real-time data integration and analysis.
MarineTraffic (marinetraffic.com)
%~\cite{MarineTraffic} 
is a web-based platform that creates a global network of vessel tracking information, enabling users to track vessels, monitor maritime traffic, analyze trends, and make informed decisions. 
It relies on the collective contributions of the maritime community, who voluntarily transmit real-time information about vessel positions and movements from various sources, including AIS (automatic identification system) data and satellite data.
MarineTraffic employs neural network architectures for automatic maritime object detection using satellite imagery~\cite{bereta2020automatic}. 
%~\cite{marinetraffic_research}. 

Waze (www.waze.com)
is a GPS navigation software that works on smartphones and other computers. 
It analyzes crowdsourced GPS data and user-submitted information along the route in real-time and uses community editing to ensure the accuracy of the map data. 
Waze uses AI algorithms to predict traffic, optimize routes~\cite{waze_medium}, and provide personalized recommendations by analyzing driver behaviors. 
Waze has implemented integration with Google Assistant, enabling users to use the application by utilizing the voice command ``Hey Google". 

By improving efficiency, reducing waste, and enabling more equitable distribution and accessibility of resources and spaces, AI can play a significant role in addressing inequalities in access to resources, paving the way for sustainable and efficient infrastructure development~\cite{cowls2021definition}.

\subsection*{Agriculture} 
Finally, the agriculture sector leverages AI-enhanced collective intelligence to improve farming practices, resource management, and crop yields, addressing global food security challenges.
WeFarm (wefarm.co)
%~\cite{WeFarm} 
is a social networking platform connecting the small-scale farming community.
It helps millions of African farmers meet, exchange solutions to their questions, and trade equipment and supplies. 
Farmers ask each other questions about agriculture and promptly receive content and ideas from other farmers worldwide through crowdsourcing.
Natural language processing processes and understands the text-based messages farmers send in different regions. 
Machine learning matching algorithms consider farmers' needs and expertise, identify the most suitable responses, and present them to the farmer who posed the question.

Mercy Corps AgriFin (www.mercycorpsagrifin.org) harnesses the power of AI and a global network of partners to transform agriculture for smallholder farmers. By utilizing state-of-the-art imagery, modeling, and analysis, AI enables precise management of crops and adaptation strategies against climate change impacts. AgriFin's integration of digital technology and data, supported by a global network, empowers these farmers in a digitally interconnected world.

In light of the global food crisis and the quest for sustainable solutions, the importance of AI-enhanced collective intelligence platforms becomes increasingly critical. These platforms can significantly contribute to addressing agricultural challenges and transitioning towards global sustainable food systems~\cite{marvin2022digitalisation}.

\subsection*{Limitations}

The previous examples across various application domains demonstrate that CI and AI leverage the power of crowdsourced data, collaborative problem-solving, and advanced data processing techniques to enhance decision-making, optimize processes, and improve outcomes. However, there are limitations and concerns associated with the AI-CI approach.

\paragraph{Scalability}
AI-enhanced collective intelligence can scale up problem-solving efficiency for real-world challenges, but scalability remains a limitation. As the system grows, scaling the multilayer framework and ensuring its applicability across various domains and scales can be challenging.
Integrating diverse fields provides a comprehensive approach but poses difficulties in coordination, consistency, and computational resources.

\paragraph{Bias}
While AI can help mitigate certain human biases, they also have the possibility to introduce new ones. These biases can stem from various stages of AI development and deployment. AI systems learn from the data they are trained on, and biased training data can cause the AI to replicate these biases. Additionally, algorithmic design choices may inadvertently favor certain groups over others. Human-in-the-loop bias is another concern, where human inputs during training and feedback, such as biased labeling by annotators can be learned by the AI. Moreover, user interactions with AI systems can also influence and reinforce these biases. AI biases pose risks to fair and equitable outcomes in society, especially in sensitive areas such as healthcare and education. Bias in AI algorithms can lead to unfair treatment or misdiagnosis in healthcare, and biased educational tools can perpetuate existing inequalities.
Understanding how AI biases and human biases interact, and how to avoid doubly biased decisions made by human-machine intelligence~\cite{silberg2019notes}, are crucial areas for further study.

\paragraph{Explainability}
%red
Many AI models, particularly deep learning algorithms, operate as ``black boxes" \cite{buhrmester2021analysis}, making understanding how they arrive at specific decisions or recommendations difficult. This opacity can hinder trust and acceptance among users, especially in high-stakes domains like healthcare and law. Explainability is essential for ensuring accountability, enabling users to understand and challenge AI decisions, and identifying and correcting biases within the system. Developing AI models that are both effective and interpretable requires ongoing research and innovation.

\section*{Challenges and outlooks}

The application areas of AI-enhanced collective intelligence continually expand as technology advances and new opportunities emerge. As introduced in the previous section, successful examples of AI-enhanced collective intelligence exist in various domains. 
However, challenges coexist with opportunities. 
When realizing human-AI collective intelligence, many aspects must be considered, such as human-machine communication, trust, crowd retention, technology design, and ethical issues \cite{tsvetkova2017understanding}. Here, we briefly discuss some of these challenges and call for further work on each aspect.

\subsection*{Communication} 

Clear and effective communication between humans and AI is the engine for collective intelligence to emerge~\cite{gupta2023fostering}.
Communication refers to the process of exchanging information between teammates~\cite{salas2008wisdom}. It is essential for team performance as it contributes to the development and maintenance of shared mental models and the successful execution of necessary team processes~\cite{salas2008wisdom, stowers2021improving}.

For effective communication in human-AI teamwork, AI must first be able to model human information comprehension accurately and then effectively communicate in a manner understandable to humans. Current AI technologies face limitations in verbal and contextual understanding~\cite{stowers2021improving} and cognitive capabilities, such as reasoning about others' mental states and intentions~\cite{matthews2021super}. 
Detection, interpretation, and reasoning about social cues from a human perspective is imperative to ensure effective coordination, but AI has yet to achieve this. Although recent developments in LLMs appear promising, further advancing our understanding of how LLMs can be utilized in such a way is crucial~\cite{lu2023emergent, schaeffer2024emergent}.

AI's inability to interpret nonverbal cues and limited self-explanation hinder human collaboration. 
Addressing these challenges, research suggests the development of AI agents with cognitive architectures that can facilitate both a Machine Theory of the Human Mind (MToHM) and a Human Theory of the Machine Mind (HToMM) will be especially important for supporting the emergence of collective intelligence~\cite{gupta2023fostering}.

\subsection*{Trust}

In team success, a human member's trust in a machine is pivotal~\cite{nam2020trust}. 
Research indicates that users interact differently with AI than humans, often exhibiting more openness and self-disclosure with human partners~\cite{mou2017media}. 
Developing explanatory methods that foster appropriate trust in AI is challenging yet crucial for enhancing performance in human-AI teams~\cite{bansal2021does}. 
Meanwhile, over-reliance on AI, especially when faced with incorrect AI advice, can be detrimental and lead to human skill degradation. Hence, it is essential for humans to judiciously rely on AI when appropriate and exercise self-reliance in the face of inaccurate AI guidance~\cite{schemmer2022should}, especially in high-stakes situations.
More tech literacy, public understanding of AI, and advancements in the technology itself will help in overcoming these challenges. 

\subsection*{Crowd retention} 

Albeit beneficial in terms of efficiency, employing AI and citizen scientists together has been reported to damage the retention of citizen scientists~\cite{trouille2019citizen}. 
There is a trade-off between efficiency and retention of the crowds when deploying AI in citizen science projects, and ultimately, if all the crowds leave the projects, the performance of AI alone would decline again.
One of the factors contributing to the retention of volunteers is reported to be the social ties they make in their teams~\cite{hristova2013life}. 
The challenge of forming social bonds with AI could increase human feelings of loneliness and isolation, potentially affecting mental health. Questions about identity and recognition could arise when humans are rewarded in conjunction with AI. 
Studies highlight concerns regarding the anthropomorphic appearance of social robots potentially threatening human distinctiveness~\cite{ferrari2016blurring}. Additionally, collaboration with AI may demotivate humans due to a lack of competitive drive and an over-reliance on AI, which could diminish human participation and initiative. Understanding how to balance AI collaboration with maintaining human motivation and retention is crucial.

\subsection*{Technology design} 

The interface design of AI technology influences user experiences, engagement, and acceptance of AI-driven communication. 
Interfaces should be user-friendly, promoting seamless information exchange and aligning with human cognitive processes.
Since AI systems need time to process human input and generate output, there may be a significant time lag between the interactions in some instances. 
The delay of feedback between human input and AI responses can adversely affect~\cite{mackenzie1993lag} the coordination and interaction efficiency between humans and AI, especially in collaborative tasks. On the other hand, human behavior possesses specific temporal patterns, mainly determined by circadian rhythm \cite{yasseri2012circadian}. 
It is essential to design such hybrid systems so that the temporal characteristics of both entities match, support mutual understanding, and balance the cognitive load and workload distribution. 
Moreover, these systems should be designed to augment human contributions~\cite{jarrahi2018artificial}. 
To effectively cater to diverse user groups, human-centered AI design approaches, including the use of personas~\cite{holzinger2022personas}, can be employed to develop adaptive and inclusive human-AI interfaces.

\subsection*{Ethical considerations}

When integrating AI into human interactions, several ethical considerations emerge. Addressing these concerns involves balancing benefits and minimizing potential harms. Key issues include the ethical implications of mandatory AI use, especially when humans are reluctant or do not require AI assistance. 
When it comes to AI transparency, a study reported that humans who were paired with bots to play a repeated prisoner’s dilemma game outperformed human-human teams initially but later started to defect once they became aware that they were playing with a bot, showing a trade-off between efficiency and transparency~\cite{crandall2018cooperating}.
Another critical aspect is addressing biases inherent in AI outputs, which are influenced by the data inputs. It is essential to consider how AI integration might exacerbate or introduce new biases in decision-making~\cite{silberg2019notes}. Furthermore, liability questions arise when AI errors cause serious harm, an issue compounded by the lack of comprehensive regulations and laws. For instance, a robot failed in object recognition and crushed a human worker to death in a factory in South Korea~\cite{bbc2023southkorea}. This highlights the urgent need for comprehensive legal frameworks to address such liabilities and ensure safety in environments where AI interacts with humans.
As AI becomes more prevalent in the workplace, constructing ethical human-AI teams becomes a complex, evolving challenge~\cite{flathmann2021modeling}.

%\subsection{Limitations}

\subsection*{Conclusion} 

The advancement of AI technologies has led to enhanced and evolved approaches to tackling existing challenges. Approaches relying solely on human or artificial intelligence encounter significant limitations. Therefore, a synergistic combination of both, complementing each other, is considered the ideal strategy for effective problem-solving.

In current real-world AI-CI applications, AI is predominantly utilized as a technical tool to facilitate data analysis through techniques such as natural language processing and computer vision. Implementing AI not only scales up tasks but, in combination with human intelligence, leads to superior performance. Although AI is critical in some decision-making processes, humans make the ultimate decisions and contributions. It is essential to recognize that AI’s function is to support and enhance human collaborative processes rather than to replace human intelligence. Nevertheless, it is also worth considering collective intelligence emerging from a group of AI entities in future research. Past work has demonstrated that the collective behavior of autonomous machines can lead to the emergence of complex social behavior \cite{tsvetkova2017even}.

AI-CI approaches present numerous challenges. Success in this field relies on multiple factors beyond the initial idea and readiness of technological solutions. Critical aspects include attracting the audience, designing user-friendly interfaces, scalability, and effectively integrating AI with human collective intelligence. Understanding how to combine human intelligence and AI to address social challenges remains a significant and worthy area of study that requires interdisciplinary collaboration.

While interdisciplinary collaboration presents challenges, such as differences in terminology, methodologies, and objectives across fields, which can create communication barriers and slow progress, it also offers benefits. Interdisciplinary collaboration fosters a comprehensive understanding of complex problems and enables developing more effective, adaptable, and ethical AI-enhanced collective intelligence systems.

Complex system thinking has illuminated various biological, physical, and social domains. Based on this, we can model real-world collective intelligence systems as multilayer networks. This framework allows us to leverage extensive research on multilayer networks, focusing on their robustness, adaptivity, scalability, resilience, and interoperability. 

Our theoretical model of human-AI collective intelligence, a multilayer network with cognition, physical, and information layers, offers valuable insights into complex phenomena in such hybrid systems. For example, in the case of the Flash Crash 2012 \cite{kirilenko2017flash}, the cognition layer includes high-frequency traders (algorithms) responding to market signals, the physical layer involves the actual financial transactions, and the information layer encompasses the data streams and communication networks connecting these elements. By understanding the interplay and feedback loops within these layers, our model helps identify vulnerabilities, such as how a single large sell order propagated through the network, causing a cascade of automated responses that led to the crash \cite{tsvetkova2024human}. 

Similarly, regarding the reported issue of human user retention while AI is being deployed in citizen science projects \cite{trouille2019citizen}, our model elucidates the interaction between human volunteers and AI classifiers within the different layers. It highlights the importance of maintaining a balance between efficiency (enhanced by AI) and volunteer engagement (rooted in human cognition). This approach prepares us to preemptively address potential pitfalls, ensuring more resilient and effective collective intelligence systems in the future.

\subsubsection*{Future research directions}
Future research in this field will require strong interdisciplinary communication between different domains and research fields. Here are a few directions for future research:

\paragraph{Behavioral Studies}
Various behavioral aspects, including human behavior, AI behavior, and the interactions between humans and AI, must be studied. To better understand and harness human-AI collectives for addressing societal challenges, researchers suggest building the foundations of Social AI~\cite{pedreschi2023social} by integrating insights from Complex Systems, Network Science, and AI.
\paragraph{Adaptive Systems}
Adaptive systems can learn and adapt over time. Researchers propose studying collective adaptation~\cite{galesic2023beyond}, focusing on how human collectives adjust their cognitive strategies and social networks in response to changing problems. Extending this research to include both human and AI collective adaptation could be valuable, requiring collaboration from cognitive science, social psychology, and other related fields.
\paragraph{Ethical guidelines}
Developing ethical guidelines and frameworks is vital to ensuring the responsible and equitable use of AI-enhanced systems and mitigating potential risks. This includes studying ethics and bias and examining broader societal implications, such as effects on employment, privacy, and social equity.
\paragraph{}
Future research should focus on large-scale empirical studies to validate theoretical models, understand real-world implications, and refine the deployment of these systems. Interdisciplinary collaborations are crucial for advancing the understanding and development of AI-enhanced collective intelligence systems. We agree that researchers across all relevant disciplines should collaborate and keep up with the latest developments to foster a promising future of AI-enhanced collective intelligence.


\subsection*{Resource availability}

%At the start of the experimental procedures section, there should be a ``resource availability" subsection, which must contain the following required subsections: \textbf{lead contact, materials availability, and data and code availability}. These sections are mandatory even if no unique reagents were generated in the study. No other subheadings or text are allowed in this section.
%\subsubsection*{Lead contact}
%\subsubsection*{Materials availability}
\subsubsection*{Data and code availability}
%    Examples: 
Our source data and code are available publicly\cite{database}.

% and has been archived at Zenodo\cite{georgios_rizos_2023_10253149}.

%\subsection*{Other experimental procedures subheadings}

%``Experimental procedures" is a Level 1 heading. ``Resource availability" is the only required Level 2 subheading. Level 3 subheadings are encouraged, and Level 4 subheadings are permitted. (5 is right out.)

\section*{Supplemental information}

%Supplemental information should be provided as separate files. In the main text, please list the files to be included, e.g.:

\begin{description}
%  \item Figures S1-S5 and their legends in a PDF
   \item {Table S1. Main concepts of key terms in human-AI collective intelligence.}

  \item {Table S2. Summary of CI and AI aspects of various application examples.}
%  \item Table S2. Another descriptive title for a different Excel file
\end{description}

\section*{Acknowledgments}

The research conducted in this publication was funded by the Irish Research Council 
under grant number IRCLA/2022/3217, ANNETTE (Artificial Intelligence Enhanced Collective Intelligence).
We thank Gianni Giacomelli for useful suggestions and the Supermind Design Augmented Collective Intelligence Database. We thank Siobhan Grayson for the valuable comments on the manuscript. 

\section*{Author contributions}

HC and TY conceived the idea. HC collected the data and performed the analysis. Both authors drafted the manuscript, read and approved the final manuscript.

\section*{Declaration of interests}

The authors declare no competing interests.

%\section*{Figures, figure titles, and figure legends}

%Figures should be cited at appropriate locations in the main text as follows (Figures 1A and 1B). Titles and legends should appear at the end of the main text (this document). For each figure, provide a brief a brief title that describes the entire figure, followed by a figure legend describing each panel (if applicable).

%\section*{Tables, table titles, and table legends}

%Tables should be numbered like figures (Table 1, Table 2, not Table 1a, Table 1b ...) and include a title. Table legends are optional but encouraged. Wherever possible, please use the Microsoft Word table function to make tables. Footnotes (with superscript lowercase letters) should be used where necessary to indicate some feature of the data; please do not use bold, italic, colored text, or shading for this purpose. Use separate cells, not line breaks or spaces, for all discrete data elements. Embedded graphics with color are OK. 

\bibliography{references}

%\paragraph{Please include only articles that are published (online publication and preprint servers are OK). Unpublished data, submitted and/or accepted manuscripts, abstracts, and personal communications should be cited within the text only (``unpublished data," ``data not shown," ``Alice Smith, personal communication") and not included in the references list. Personal communication should be documented by a letter of permission. Whenever possible, please make sure your .bib file has complete author lists for each item.}

\section*{MAIN FIGURE TITLES AND LEGENDS}
%\noindent\includegraphics[width=0.85\linewidth]{new_multiplex.pdf}

\begin{figure}[htbp]
    \begin{center}
      \includegraphics[scale=0.9]{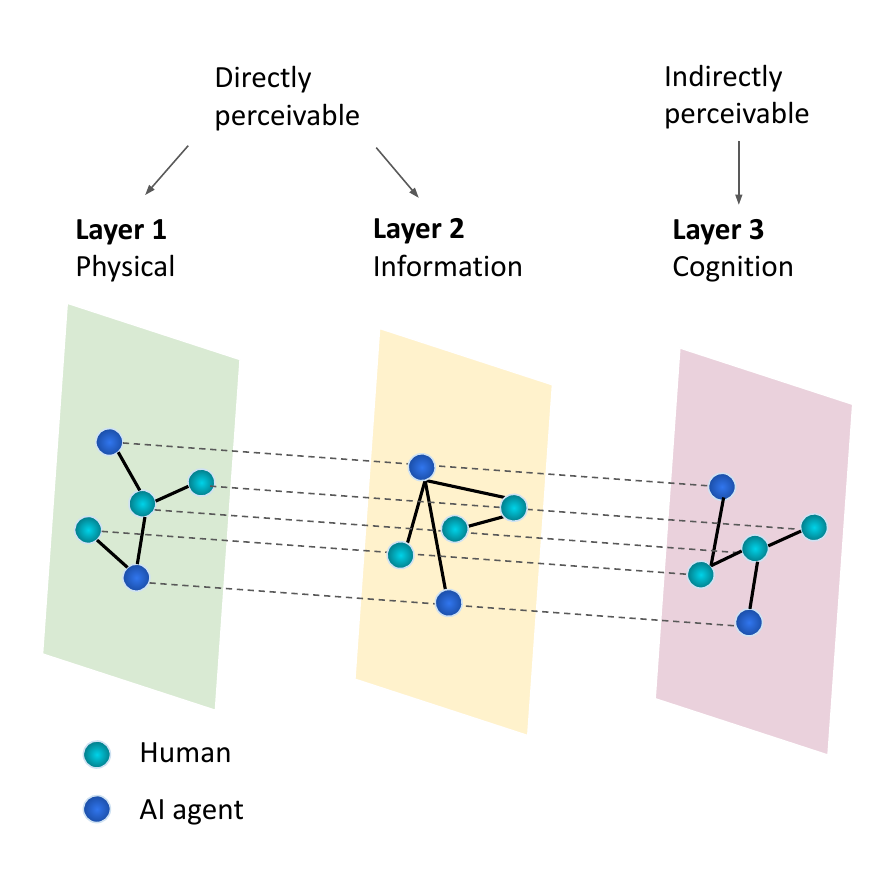} %0.08
    \end{center}
    \caption{{\bf A multilayer representation of a collective intelligence system.} A multilayer representation can be used to untangle the processes within the complex system of human and AI agents. It consists of three interdependent layers: cognition, physical, and information. External factors and a changing environment can also influence the entire system's emergent collective intelligence.} 
    \label{fig:fig1}
\end{figure}

\begin{figure}[htbp]
    \begin{center}
      \includegraphics[scale=0.7]{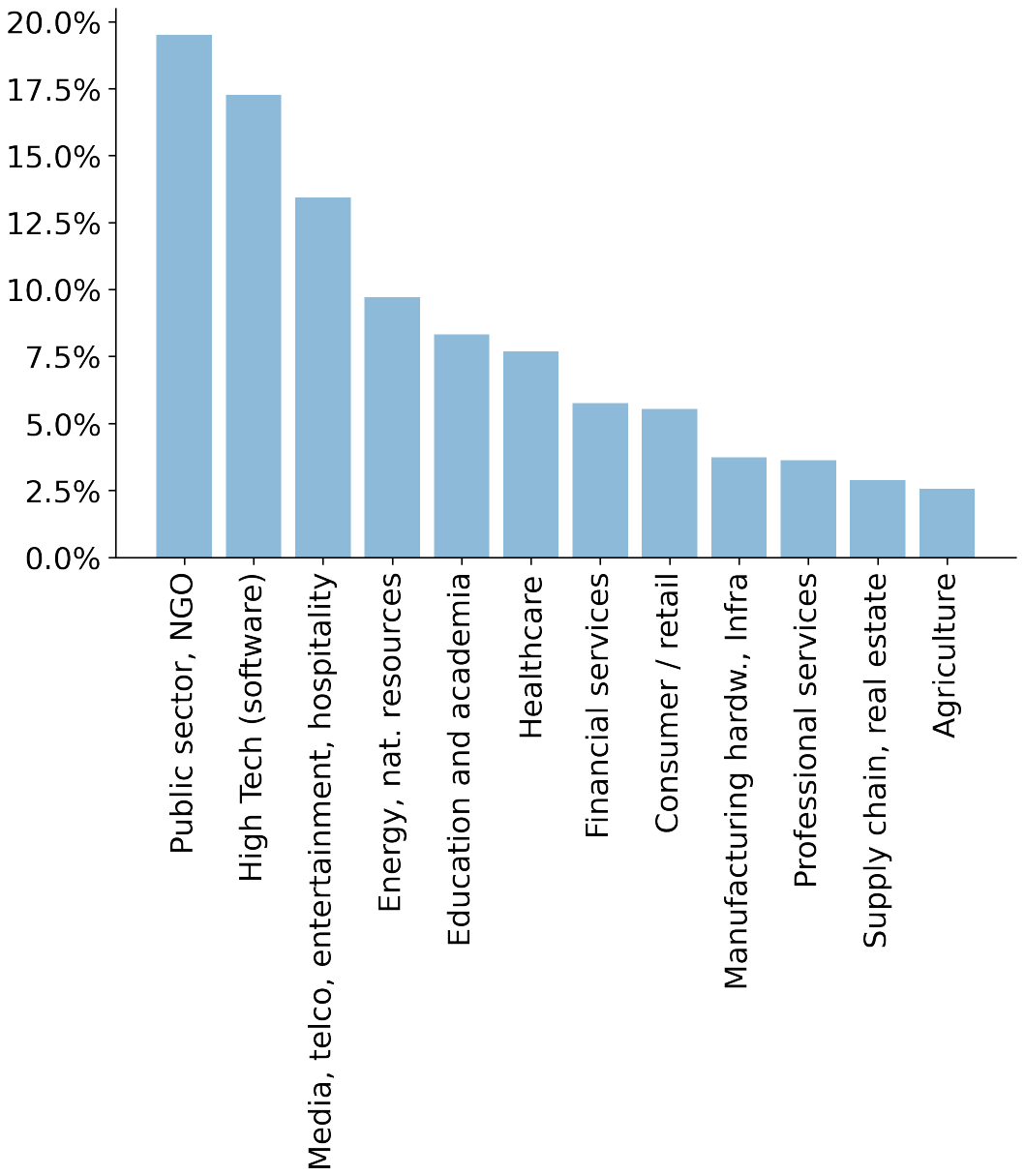} 
    \end{center}
    \caption{{\bf AI-CI cases by application area}. Distribution of AI-enhanced collective intelligence cases by application area based on dataset curated by Supermind Design.} 
    \label{fig:fig2}
\end{figure}

%\section*{Additional manuscript components}

%Depending on the article type, you may be asked to upload the following as separate files: graphical abstract, Bigger Picture, highlights, and eTOC (In Brief). These items are not required for initial submissions. 

%Patterns may require additional manuscript elements to be submitted with the final files; please refer to the journal's homepage and Final Files Checklist for details.

\newpage



\section*{Supplementary material}

| Term | Main concepts | References |
|---|---|---|
| AI-enhanced Collective Intelligence | A system where artificial intelligence tools and techniques are integrated to augment the collective intelligence of a group, improving decision-making, problem-solving, and innovation. | |
| Complex Adaptive System | A subset of complex systems characterized by the ability of its components to adapt and learn from interactions with each other and the environment, leading to dynamic evolution. | [59] |
| Collective Intelligence | The group intelligence that emerges from the collaboration and collective efforts of individuals working together towards a common goal. | [2] |
| Complex System | A system composed of interconnected parts that exhibit collective behavior, self-organization, and adaptation to changing environments, often displaying emergent properties. | [48] |
| Complexity Theory | A framework for understanding how interactions among components of a system give rise to collective behaviors and emergent phenomena that cannot be predicted from individual parts. | [48] |
| Emergence | The process through which larger patterns, structures, or behaviors arise from the interactions among smaller or simpler entities that do not exhibit such properties individually. | [59] |
| Hybrid Intelligence | The combination of human intelligence and artificial intelligence, leveraging the strengths of both to solve complex problems more effectively. | [32] |
| Multilayer Network | A network model where nodes represent entities and different layers represent different types of interactions or relationships between these entities. | [55] |
| Wisdom of Crowds | The collective opinion of a diverse and independent group of individuals can be more accurate than that of a single expert, leveraging diversity of thought and decentralization. | [10] |

Table S1: Main concepts of key terms in human-AI collective intelligence. For references, see the main article.

| Application | Area | Aspects of Collective Intelligence | Aspects of Artificial Intelligence |
| --- | --- | --- | --- |
| Safecity | Public sector, NGO | <ul><li>Crowdsourced reporting</li><li>Community discussion and initiatives</li></ul> | <ul><li>Predictive analytics</li><li>Sentiment analysis</li></ul> |
| Ushahidi | Public sector, NGO | <ul><li>Crowdsourced crisis information</li><li>Community reporting</li></ul> | <ul><li>Machine learning</li><li>Natural language processing</li></ul> |
| Bluenove | High Tech (software) | <ul><li>Open consultation</li><li>Collaborative ideation</li></ul> | <ul><li>Natural Language Processing</li><li>Emotion analysis</li></ul> |
| Figure Eight | High Tech (software) | <ul><li>Crowdsourced annotation</li></ul> | <ul><li>Machine learning</li><li>Automated Workflows</li></ul> |
| Civil War Photo Sleuth | Media, telecommunication, entertainment, hospitality | <ul><li>Crowdsourced Identification</li><li>Collaborative research</li></ul> | <ul><li>Facial recognition</li><li>Image analysis</li></ul> |
| Bellingcat | Media, telecommunication, entertainment, hospitality | <ul><li>Crowdsourced investigation</li><li>Community collaboration</li></ul> | <ul><li>Multi-modal data analysis</li><li>Pattern recognition</li></ul> |
| Litterati | Energy, natural resources | <ul><li>Crowdsourced litter data</li><li>Community engagement</li></ul> | <ul><li>Computer vision</li><li>Predictive modeling</li></ul> |
| eBird | Energy, natural resources | <ul><li>Crowdsourced bird observations</li><li>Peer review and validation</li></ul> | <ul><li>Computer vision</li><li>Predictive modeling</li></ul> |
| Kialo Edu | Education and academia | <ul><li>Structured debates and discussions</li></ul> | <ul><li>Natural language processing</li><li>Sentiment analysis</li><li>content recommendation</li></ul> |
| Zooniverse | Education and academia | <ul><li>Crowdsourced research</li><li>Collaborative data analysis</li></ul> | <ul><li>Machine learning classification</li><li>Pattern recognition</li></ul> |
| Human Dx | Healthcare | <ul><li>Crowdsourced medical expertise</li><li>Collaborative problem-solving</li></ul> | <ul><li>Pattern recognition</li><li>Data analytics</li></ul> |
| CrowdEEG | Healthcare | <ul><li>Collaborative annotation</li></ul> | <ul><li>Machine learning training</li><li>Predictive modeling</li></ul> |
| Numerai | Financial services | <ul><li>Crowdsourced predictions</li><li>Community collaboration</li></ul> | <ul><li>Ensemble Learning</li><li>Predictive Analytics</li></ul> |
| CryptoSwarm AI | Financial services | <ul><li>Crowdsourced market insights</li><li>Swarm intelligence</li><li>Community feedback</li></ul> | <ul><li>Swarm AI</li><li>Predictive modeling</li><li>Automated alerts</li></ul> |

| MarineTraffic | Supply Chain, Real Estate | ● Crowdsourced Vessel Tracking<br>● Community Contributions<br>● Data Sharing | ● Automatic Identification System (AIS) Data Analysis<br>● Object detection<br>● Predictive Analytics |
|---|---|---|---|
| Waze | Supply Chain, Real Estate | ● Crowdsourced Traffic Data<br>● Community Editing and feedback | ● Route Optimization<br>● Personalized Recommendations<br>● Voice assistant integration<br>● Predictive Analytics |
| WeFarm | Agriculture | ● Crowdsourcing advice<br>● Knowledge sharing<br>● Problem-solving | ● Automated matching<br>● Personalized recommendations<br>● Natural language processing |
| Mercy Corps AgriFin | Agriculture | ● Community-driven peer learning<br>● Knowledge sharing | ● AI imagery<br>● Data analytics and modeling<br>● Recommendation algorithm |

Table S2: Summary of CI and AI aspects of various application examples.



\end{document}